\documentclass[aps,pra,superscriptaddress,amsmath,floatfix,twocolumn]{revtex4}
\usepackage{graphicx}  
\usepackage{subfigure}
 \usepackage{epsfig} 
 \usepackage{bm}
\begin{document}
\def\bar{\begin{eqnarray}}
\def\ear{\end{eqnarray}}
\def\beq{\begin{equation}}
\def\eeq{\end{equation}}
\title{Resonant Transmission of Electromagnetic Fields
Through Subwavelength Zero-$\epsilon$ Slits}
\author{Klaus Halterman}
\author{Simin Feng}
\affiliation{Physics Division, Naval Air Warfare Center,
China Lake, California 93555}
\date{\today}


\begin{abstract}
We theoretically investigate the transmission of electromagnetic radiation through 
a metal plate with a
zero-$\epsilon$ metamaterial slit,
where the permittivity tends towards zero
over a given bandwidth. 
Our analytic results demonstrate that 
the transmission
coefficient can be substantial for a broad
range of slit geometries, including
subwavelength widths that are many wavelengths long. 
This novel resonant effect has features 
quite unlike the Fabry-P\'{e}rot-like resonances that
have been observed in conductors with deep channels.
We further reveal  that 
ultranarrow zero-$\epsilon$ channels
can have significantly {\it greater} transmission compared
to slits with no wave impedance difference across them. 
\end{abstract}
\maketitle
With the current state of the art in nanofabrication technologies,
and the recent
observation of resonant optical transmission
through small metal holes \cite{ebb}, 
there has been a renewed interest in electromagnetic wave transmission
and diffraction
in metallic nanostructures. Although diffraction effects
involving optical slits and gratings has a long history,
scientific curiosity coupled with the potential 
for device applications has spurred considerable research activity involving
the manipulation and confinement of
electromagnetic waves in
nanoscale resonant structures. With the concurrent advent of  
metamaterials, 
or composite structures with tailored electromagnetic
properties, the interplay of
classical transmission systems with these new composite media
has become an important issue.
By engineering metallic nano-apertures, gratings,
or channels to incorporate metamaterials, it is anticipated that
various enhanced or exotic transmission characteristics will ensue.
\newline
\indent
Lately, there has been interest in
zero-$\epsilon$ metamaterials, which are structures that exhibit 
an effective
permittivity of  zero (or nearly so)  in the passband,
akin to the property of some precious metals \cite{palik} near their plasma frequency.
It was shown  that 
waveguide devices containing zero-$\epsilon$ inclusions
could potentially be more efficient via the reduction of unwanted reflections
if one of the physical dimensions was made smaller \cite{eng}.
This can possibly negate the ill effects of impedance mismatch.
A multilayer structure having nonlinear electric and magnetic responses,
and a refractive index close to zero,
was shown to effectively shield \cite{feng} electromagnetic fields.
Experimental work has verified \cite{liu,eng3} electromagnetic tunneling through 
zero-$\epsilon$ metamaterials at microwave frequencies.
It was shown \cite{eng2} that  
the effective magnetic permeability, $\mu$,  
can also be resonantly tuned to vanish,  creating a matched metamaterial
with an effective zero index of refraction.
These media have been linked to
applications in miniaturized resonators, highly directive
antennas \cite{enoch}, delay lines with zero-phase difference
I/O, and transformers
that convert small-curvature wave fronts
into output beams with planar-like
wave fronts \cite{zio}.
\newline
\indent
A fundamental system 
in which to investigate zero-$\epsilon$ diffraction and transmission effects 
is a 
subwavelength channel through a metal film.
If the slit is 
filled with air, 
it has been established that for light polarized perpendicular to 
the slit in an optically thick perfect metal of length, $l$, 
Fabry-P\'{e}rot-like 
waveguide modes arise when $l$
is approximately a half integer number of wavelengths: $l\approx n\lambda/2$. 
These harmonic modes, which follow from geometrical arguments, 
result in transmission peaks when the waves coherently 
superimpose over the given path length.
As the geometrical parameters vary, the resonant wavelength can shift \cite{takakura,mybad}
in  metals with slit perforations, as observed with microwaves \cite{suck0}.
For ultrashort incident pulses,
the Fabry-P\'{e}rot-like modes
can be resonantly activated \cite{mechler}
through the Fourier components of the wave packet inside the slit,
leading in some cases to enhanced transmission.
The resonant enhancement of a nanometer scale electromagnetic pulse was also shown to
be spatially and temporally localized \cite{mechler2} in the near field. 
For perfect metals with holes rather than slits,
transmission 
is strictly limited to
incident
wavelengths
that are less than twice the diameter of the openings.
\newline
\indent
In this Letter we reveal some
unexpected and
exotic transmission behavior of
light through a subwavelength zero-$\epsilon$ slit. 
Our theoretical framework demonstrates
that significant transmission can occur in these structures 
for a considerable range of ultranarrow widths that are sufficiently
deep. The inverse relationship between the slit length and width that achieves  
maximal transmission is shown to be highly nontrivial. 
We show that  zero-$\epsilon$ metamaterials with inherently large
intrinsic
impedances, can have  greater transmission than
matched zero index slits, where $\mu$ is also vanishingly small.
We also take advantage of the subwavelength geometry (that is, the slit width is
smaller  than about half the incident wavelength),
to invoke the single mode approximation, which has been shown to yield valuable 
physical insight into
resonance phenomena for perfect \cite{mata} and real \cite{gordy} metals.
Our analytic results thus
permit one to efficiently map out the relevant sector of parameter space
deemed appropriate 
for any future experimental endeavors.
\newline
\indent
We consider a planar perfect metallic structure that is translationally invariant 
in the $x-z$ plane and has  channel
length $l$ normal to the plane
with width $p$ (along $x$). 
The incident beam is TM polarized, so that the magnetic field ${\bm H}$ is directed along
$z$. The wavevector, ${\bm k}$, forms an angle $\theta$ with the normal to the plane.
For this configuration, the $z$ component of the magnetic field, $H_z$, 
must satisfy the scalar Helmholtz equation,
\begin{align}
\label{helm}
\frac{\partial^2 H_z}{\partial x^2} +  \frac{\partial^2 H_z}{\partial y^2} + \epsilon \mu k_0^2 H_z=0,
\end{align}
where $k_0=\omega/c$.
Our focus is primarily  metamaterial slits very near
the characteristic plasma frequency 
and, as in recent works, represent the frequency dispersive electrical response 
by an effective Drude model,
$\epsilon=1-\omega_p^2/[\omega(\omega+ i \Gamma)]$,
where $\Gamma$ is
correlated to the mean free path in the filling material.
Other than when discussing absorption loss, 
we generally take $\Gamma$ equal to zero in order to isolate
the electromagnetic effects inherent to vanishing constitutive relations.


We solve (\ref{helm}) 
in a given region,
and then match each solution
at the corresponding interfaces
using the appropriate boundary conditions.
To construct the solution in the ``continuum" regions of free space surrounding the
metal sheet, we
Fourier transform Eq.~(\ref{helm}) along $x$ and use separation of variables.
This renders an expression written in terms of  plane wave expansions,
\begin{align}
H_{z,1}&=e^{i (k_{0x}x  - k_{0y}y)}+\frac{1}{\sqrt{2\pi}}\int_{-\infty}^{\infty} d\alpha
R(\alpha)e^{i\alpha x}e^{i\beta y} \\
H_{z,3}&=\frac{1}{\sqrt{2\pi}}\int_{-\infty}^{\infty} d\alpha
T(\alpha)e^{i\alpha x}e^{-i\beta y},
\end{align}
where $k_{0x}=k_0\sin\theta$, $k_{0y}=k_0\cos\theta$, and $\beta=\sqrt{k_0^2-\alpha^2}$.
Here the subscripts 1 and 3 denote the entrance and exit regions respectively.
The unknown coefficients $R(\alpha)$ and $T(\alpha)$ are determined below.
The general eigenmode expansion
for the field within the slit region is \cite{takakura},
\begin{align}
\label{mode}
H_{z,2}=\sum_{m=0}^{\infty} \cos\bigl[k_m (x-p/2)\bigr]\bigl[a_m e^{i \tau_m y} + b_m e^{-i \tau_m y} \bigr],
\end{align}
where $k_m\equiv m\pi/p$, and 
$\tau_m\equiv \sqrt{\epsilon \mu k_0^2-k_m^2}$. 
The summation is over a single index $m$ reflecting the 
interconnection among the
wavenumbers consistent with the wave equation. 
The 
electric field is obtained directly via 
application of Maxwell's equation: ${\bm E}=i/(\epsilon k_0)\nabla\times {\bm H}$.
It is generally valid to retain only the lowest order
mode in the expansion (\ref{mode})
for the ultranarrow channels considered in this paper.
Moreover, as $\epsilon$ goes to zero, 
the summation results in
higher order evanescent fields
that have a negligible contribution to the time-averaged energy flux.
Thus, for the electric field we have,
\begin{align}
E_{x,2}=\eta(-a e^{i \tau y} + b e^{-i \tau y})\zeta(x), \label{ex2}
\end{align}
where  $\tau\equiv\tau_0$, 
$\eta = \sqrt{\mu /\epsilon }$ is the intrinsic impedance,
and the function $\zeta(x)$
is unity in  the slit and vanishes elsewhere. 

Matching the electric field at the entrance and exit interfaces, yields the $R$
and $T$ coefficients:
\begin{subequations}
\begin{align}
R(\alpha)&= \frac{\sqrt{2\pi} k_{0y} \delta(k_{0x}-\alpha)}{\beta}+
\frac{w}{\beta} (a-b) \Phi(\alpha p), \label{balph}  \\
T(\alpha)&=\frac{w}{\beta} (-a e^{-i \tau l}+ b e^{i\tau l})e^{-i \beta l} \Phi(\alpha p), 
\end{align}
\end{subequations}
where we define $\Phi(x) \equiv \sqrt{2/\pi} \sin(x/2)/x$. 
A convenient technique \cite{mata} that utilizes the
 boundary conditions to determine the remaining unknown 
$a$ and $b$ coefficients is the use of the following relationships 
at the appropriate openings:
$\langle H_{z,1} (x,0),\zeta(x) \rangle = \langle H_{z,2}(x,0),\zeta(x) \rangle$,
and $\langle H_{z,2} (x,-l),\zeta(x) \rangle = \langle H_{z,3}(x,-l),\zeta(x) \rangle$.
After some tedious algebra, we have for $b$, 
\begin{align}
b=\frac{2\sqrt{2\pi}\Phi(k_{0x} p) (1+w I_0)}{(1+w I_0)^2-e^{2 i \xi}(1-w I_0)^2},
\end{align}
where $w\equiv k_0 p \eta$, $\xi\equiv \tau l$, and 
$a=b e^{2 i \xi}(w I_0  - 1)/(w I_0+  1)$.
The complex valued integral, $I_0$ is a function solely of $q$:
\begin{align}
I_0 &= \frac{1}{\pi q^2}\int_{0}^\infty d u \frac{\sin^2(u q)}{u^2\sqrt{1-u^2}},
\label{I3}
\end{align}
where $q\equiv k_0p/2$ is the dimensionless measure characterizing the slit width.
The series expansion for the imaginary component of $I_0$
has the following leading terms: 
\begin{align}
{\rm Im}[I_0] \approx -\frac{1}{6\pi}\Bigl[ ( q^2-6)
(\gamma +\ln q ) + 9 -\frac{19}{12}
q^2\Bigr],
\label{I0a}
\end{align}
where $\gamma$ is Euler's constant.
Similarly for the real part of $I_0$, we have (to fourth order),
${\rm Re}[I_0] \approx
\frac{1}{2} - \frac{1}{12}q^2 + \frac{1}{120}q^4$.
It will be seen  shortly that $I_0$
is a very important quantity in the determination of the transmission.
\begin{figure}
\centering
\includegraphics[width=.47\textwidth]{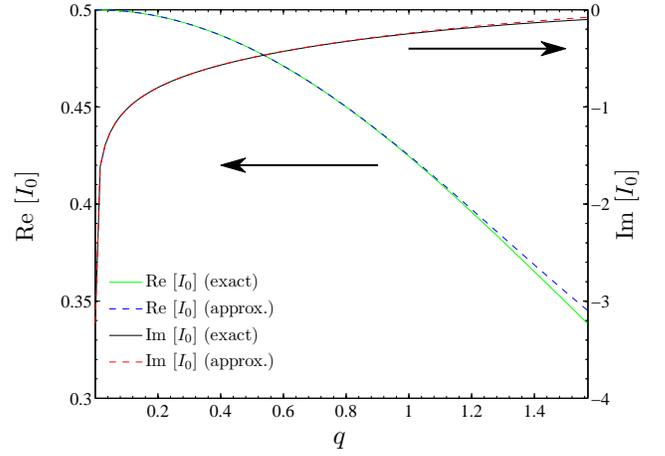}
\caption{Real and imaginary components of $I_0$ as
a function of the dimensionless width $q\equiv \pi p/\lambda$. The 
approximate solutions and exact integral [Eq.~(\ref{I3})]
have good agreement over the relevant range of widths.
As $q$ vanishes,
the imaginary component has a slow divergence [see Eq.~(\ref{I0a})].
}
\label{I3check} 
\end{figure}
The accuracy of these expansions is demonstrated graphically in
Fig.~\ref{I3check}, where 
the truncated expansions are plotted alongside
the numerically integrated $I_0$ [Eq.~(\ref{I3})].
Clearly, the approximations for $I_0$ are satisfactory
for the small $q$ of interest here, deviating only slightly for 
the larger $q$, as would be expected for a power series centered about the origin.

Next, we insert the calculated 
$a$ and $b$ coefficients into Eq.~(\ref{ex2}),
to determine  the
electric field at the bottom of the slit,
\begin{align}
\label{ex}
E^{\rm bot}_x=\frac{2i \sin(q\sin\theta) \eta}{q\sin\theta[\sin\xi+4 I_0 q\eta_2 (i\cos\xi+I_0 q\eta_2\sin\xi)]},
\end{align}
which is valid for arbitrary $\mu$ and $\epsilon$. 
The magnetic field is easily obtained via the relationship, $H^{\rm bot}_z = 2 q I_0 E^{\rm bot}_x$.
The 
transmission coefficient, 
$\tau$,
is defined as the time-averaged 
Poynting vector
at the bottom of the slit, $\langle S_y \rangle_{y=-l}$,
integrated over the exit opening, and divided by 
the  
Poynting vector of the incident plane wave: 
\begin{align}
\label{T2}
\tau=\frac{16 q^2 {\rm Re}[I_0] |\eta|^2 {\cal F}(\theta)}{|\sin\xi+ 4 I_0  q \eta (i\cos\xi+I_0  q 
\eta\sin\xi)|^2},
\end{align}
which is dimensionless after incorporating a normalization factor, $k_0$. The 
angular dependence is encapsulated entirely by the expression,
${\cal F}(\theta)=
  {\rm sinc}^2(q\sin\theta)/\cos\theta$, 
which depends weakly on $q$ 
for subwavelength slits, and is close to unity for 
source waves near normal incidence.
For small $q$, we can further write (for $\sin \xi \neq 0$):
$\tau \approx 8(\mu /\epsilon ) q^2 \csc^2(k_0l\sqrt{\epsilon \mu })$,
otherwise if we  
consider channels with  some integer multiple of the Fabry-P\'{e}rot length,
$l=\lambda/(2 \sqrt{\epsilon\mu})$,  
Eq.~(\ref{T2}) approximately reduces to,
$ \tau \approx \frac{1}{2}|I_0|^{-2}$,
which clearly only depends on the ratio of the slit width to the wavelength [see Eq.~(\ref{I3})].

Having derived the general expression for transmission of electromagnetic fields
through a 
subwavelength channel containing, to this point, conventional material parameters,
we now examine the effects of taking the limit of vanishing $\epsilon$.
It is relatively straightforward to show that Eq.~(\ref{ex}) for the electric
field at the slit exit now reduces to the  succinct form,
\begin{align}
E^{\rm bot}_x =\sqrt{\frac{\pi}{2}} \frac{\Phi(2q\sin\theta)}{ q I_0 (1-i I_0 k_0 l q \mu )}, 
\end{align}
giving a  transmission of, 
\begin{align}
\label{Te0}
\tau=\frac{{\rm Re}[I_0]}{|I_0|^2 \bigl|1-i I_0 k_0 l q \mu \bigr |^2},
\end{align}
where $H_z = 2 q I_0 E^{\rm bot}_x$, and is spatially constant in the slit.
It can be deduced from Faraday's Law, ${\bm H}=-i/(\mu k_0)\nabla\times {\bm E}$,
that $E_x$ must therefore be a {\it linear} function of the coordinate $y$ within the slit:
$E_x = (y/l) \delta E_x  + E_x^{\rm top}$, where $\delta E_x \equiv E_x^{\rm top}-E_x^{\rm bot}$.
The electric field at the top of the slit, $E_x^{\rm top}$,
is simply, $E_x^{\rm top} = (1-2 i q I_0 k_0 l\mu)E_x^{\rm bot}$.
For $q\ll1$, we can write Eq.~(\ref{Te0}) strictly in terms of elementary functions,
\begin{align}
\tau\approx 
\frac{2\pi[\pi+k_0l \mu  q (3-2\gamma-2\ln q)]}
{4\ln q(2\gamma-3+\ln q) + \pi^2 +(2\gamma-3)^2}, 
\label{trans2}
\end{align}
demonstrating that for a given ratio of slit width
to wavelength,
$\tau$
is simply a linear function of the channel length $l$ and permeability $\mu$.

\begin{figure}
\centering
\includegraphics[width=.44\textwidth]{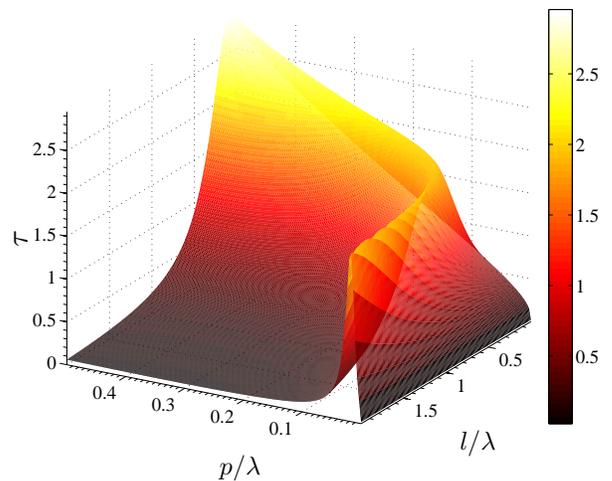}
\caption{Transmission, $\tau$, through
a zero-$\epsilon$ slit as a function of the dimensionless 
geometrical parameters, $l/\lambda$ and $p/\lambda$.
}
\label{f1} 
\end{figure}
To illustrate  how the transmission depends upon the geometry
of the slit, we present in Fig.~\ref{f1}, a three-dimensional representation
of $\tau$ [Eq.~(\ref{Te0})] as a function of the slit
dimensions, $l$ and $p$, scaled by the wavelength.
In this figure, the source field is normally incident
with wavelength, $\lambda=5\mu{\rm m}$, which is very close
to the resonant plasma wavelength,
so that the frequency dispersive $\epsilon$ nearly vanishes.
It is evident from the plot that the
transmission does not possess
Fabry-P\'{e}rot-like resonant oscillations as
a function of $l$, as would be expected from
a slit with  conventional material. 
The figure shows that energy flow is generally restricted unless
the geometrical parameters 
coincide with the brightly peaked regions that can decay quite rapidly.

\begin{figure}
\centering
\includegraphics[width=.44\textwidth]{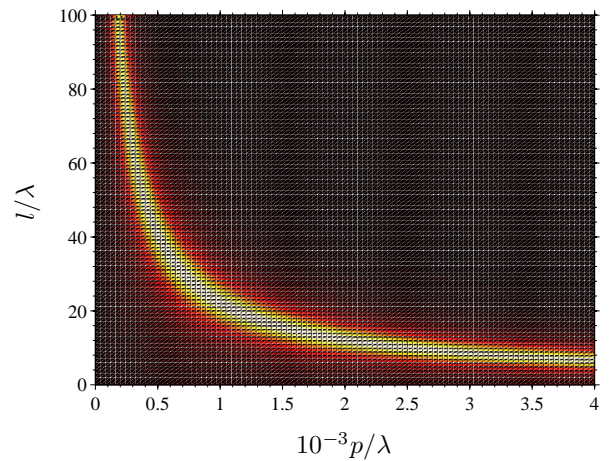}
\caption{Transmission characteristics for a
zero-$\epsilon$ metamaterial slit, demonstrating the $l$
and $p$ that give the optimal transmission at
a given wavelength. Brighter regions indicate
higher transmission. 
Relatively high transmission can persevere in deep slits ($l \gg \lambda$)
and small openings ($p \ll \lambda$). 
}
\label{f2} 
\end{figure}
\begin{figure}
\centering
\includegraphics[width=.48\textwidth]{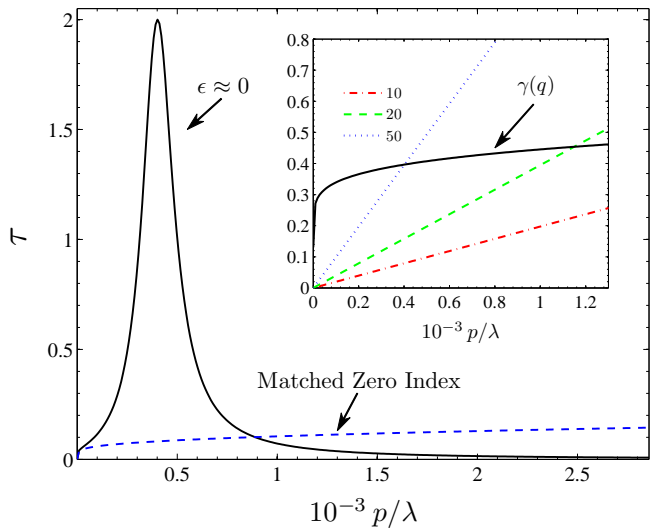}
\caption{Transmission through deep metamaterial slits as a function
of $p/\lambda$. A considerable resonance peak emerges
for a  small subwavelength width of zero-$\epsilon$ material. Also shown is 
a matched zero index medium where $\mu$ and $\epsilon$ are both nearly zero. 
The inset shows the intersections of the lines with
the $\gamma(q)$ curve, corresponding to electrical widths that yield peak transmission.
Differing ratios of $l/\lambda$ (see legend) yield lines with different slopes.
These results are
consistent with the main plot, where the channel length corresponds to $l=50\lambda$.
}
\label{f3} 
\end{figure}

The wave impedance, $Z$, defined as
$Z=E_x/H_z$,
is calculated from the field expressions above
to yield the following impedance relation between the top and  bottom of
the openings, 
$Z^{\rm top} = Z^{\rm bot} - i k_0 l \mu$,
with $Z^{\rm bot} = 1/(2 q I_0)$.
A remarkable property of a  narrow zero-$\epsilon$ slit is that
despite the considerable wave impedance
for $p \ll \lambda$, energy can still be transmitted if
the 
slit length $l$ is {\it increased}. 
This behavior is
further illustrated in Fig.~\ref{f2}, where the transmission
is shown to be robust for an extended range of channel lengths and widths.
The maximum transmission, $\tau_{\rm max}$,
seen contained within the continuous bright curve, 
is determined by examining the extrema of Eq.~(\ref{Te0}).
By taking the appropriate derivative of the denominator, we find
that $\tau_{\rm max}$ is
determined by the transcendental equation,
$\mu  q k_0 l = \gamma(q)$, where $\gamma(q)=-{\rm Im}[I_0]/|I_0|^2$
is a gradually increasing  monotonic  function of $q$.
Thus, as a function of $q$, and for each $l$,
the intercept of $\gamma(q)$ with lines of slope $\mu  k_0 l$
yields the permissible electrical slit widths
that achieve peak light transmission. 
This technique is shown graphically in the inset of Fig.~\ref{f3}, where
the opposite relationship between  slit width and length are
clearly seen.

It is also of interest to determine if  the transmission
can be  enhanced for light 
incident upon a  a  matched zero index material, whereby 
both $\epsilon$ and $\mu$ 
are vanishingly small.
In the main plot of Fig.~\ref{f3}, the transmission is therefore
plotted as a function of slit width
for a zero-$\epsilon$ slit (with $\mu=1$ as usual),
and matched zero index slit.
Surprisingly, for  zero-$\epsilon$ media, 
the shown subwavelength widths support transmission resonances that are absent 
in matched zero index slits. Although the $l$ dependence washes out
for matched zero index channels,
$\tau$ still depends strongly on the width dimension, and Eq.~(\ref{Te0})
approximately reduces to the expression
for a
slit filled with a conventional
dielectric and length satisfying the Fabry-P\'{e}rot 
geometric resonance condition.
We can thus conclude that 
for zero-$\epsilon$ media
and narrow enough openings, energy flow can be much greater than 
for slits loaded with metamaterials having no wave impedance mismatch between
the ends of the slit.
We also considered the effects of finite $\Gamma$, and concluded 
that the presence of absorption loss
simply reduces the overall magnitude of $\tau$ by a factor that becomes
greater with increasing slit depth, but having no effect
on the geometrical parameters leading to the particular resonance location.
This bodes well for future development of miniaturized devices
used for energy transport and narrow band frequency selective surfaces.

In conclusion, we have shown that significant transmission arises for 
a range of subwavelength widths
in sufficiently deep slits
containing zero-$\epsilon$ inclusions. We also demonstrated that these resonant channels can transmit
greater energy compared to the corresponding matched zero index slits.
Although there is currently limited metamaterial
fabrication in the IR and optical frequencies, 
some recent progress has been made with negative-index
metamaterials involving metal layers separated by a dielectric \cite{ref11}, 
alternating layers of InGaAs and AlInAs
semiconductors \cite{ref12}, and photonic crystals in the near-infrared range \cite{ref13}.

\acknowledgments
This work is supported by NAVAIR's ILIR program
sponsored by ONR and by a grant of HPC resources 
from  ARSC at the University of Alaska, Fairbanks,
as part of the DOD HPCMP.

\end{document}